\documentstyle[12pt,aasms4,psfig]{article}

\slugcomment{Accepted to Astrophysical Journal, 2000, Jan.}

\lefthead{Tamura et al.}
\righthead{The Potential Profile of Abell 1060}

\begin{document}

\title{X-ray Measurements of the Gravitational Potential Profile 
in the Central Region of the Abell 1060 Cluster of Galaxies}

\author{Takayuki Tamura}
\affil{Institute of Space and Astronautical Science,
			3-1-1 Yoshinodai,
			Sagamihara,
			Kanagawa  229-8510, Japan}

\author{Kazuo Makishima\altaffilmark{1}, Yasushi Fukazawa}
\affil{Department of Physics, University of Tokyo, 
7-3-1 Hongo, Bunkyo-ku, Tokyo 113, Japan}

\author{Yasushi Ikebe}
\affil{Max-Planck-Institut f\"{u}r extraterrestrische physik,
Postfach 1603, D-85740, Garching, Germany}

\and

\author{Haiguang Xu}
\affil{The Institute for Space and Astrophysics, 
Department of Applied Physics, School of Science,
Shanghai Jiao Tong University, 1954 Huashan Road,
Shanghai, 200030, PRC}

\altaffiltext{1}{also Research Center for the Early Universe (RESCUE),
University of Tokyo, 7-3-1 Hongo, Bunkyo-ku, Tokyo 113, Japan}



\begin{abstract}
X-ray spectral and imaging data from {\it ASCA} and {\it ROSAT} were used to measure the total mass profile
in the central region of Abell 1060, 
a nearby and relatively poor cluster of galaxies.
The {\it ASCA} X-ray spectra, after correcting for the spatial response of the X-ray telescope, 
show an isothermal distribution of the intra-cluster medium (ICM) 
within at least $\sim$ $12'$ (or $160h_{70}^{-1}$ kpc; $H_0 = 70\ h_{70}$ km s$^{-1}$Mpc$^{-1}$) in radius of the cluster center.
The azimuthally averaged surface brightness profile from the {\it ROSAT} PSPC
exhibits a central excess above an isothermal $\beta$ model.
The ring-sorted {\it ASCA} GIS spectra and the radial surface brightness distribution from the {\it ROSAT} PSPC were simultaneously utilized to constrain the gravitational potential profile.
Some analytic models of the total mass density profile were examined.
The ICM density profile was also specified by analytic forms.
The ICM temperature distribution was constrained to satisfy the hydrostatic equilibrium, 
and to be consistent with the data.
Then, the total mass distribution was found to be described better by the universal dark halo profile proposed by Navarro, Frenk, and White (1996;1997) 
than by a King-type model with a flat density core.
A profile with a central cusp together with a logarithmic radial slope of $\sim 1.5$ was also consistent with the data.
Discussions are made concerning the estimated dark matter distribution around the cluster center.
\end{abstract}




%

\section{INTRODUCTION}\label{sect:intro}
Measurements of the gravitational potential in clusters of galaxies provide one of the best ways to investigate the large-scale distribution of dark matter, 
and to constrain models of cosmic structure evolution.

The potential structures of clusters have been probed mainly by X-ray observations of the intra-cluster medium (ICM).
In previous studies,
ICM density distributions were usually represented by a $\beta$ model, 
which approximates an isothermal ICM hydrostatically confined in a King-type (King 1962) potential with a flat density core
(e.g. Jones and Forman 1984).
However, the King potential , or its modifications, 
are not the only possible solutions to the equation describing isothermal self-gravitating systems, 
and there is no a priori reason to believe that this particular type of potential is realized in actual clusters.

On the basis of N-body simulations of cold dark matter particles,
Navarro, Frenk, \& White (1996, 1997; hereafter NFW96 and NFW97 respectively) have shown that the density profiles of the simulated mass clumps (halos) can be 
universally described by a simple analytic formula as
\begin{eqnarray}
\rho _{\rm tot}^{\rm NFW} (r) & \propto & \left(\frac{r}{r_{\rm s}}\right)^{-1} \left(1+\frac{r}{r_{\rm s}}\right)^{-2}, \label{eq:nfw-dm}
\end{eqnarray}
where $r$ is the three dimensional radius and $r_s$ is a scale radius.
This profile, hereafter referred to as the NFW profile, 
exhibits a singularity cusp at the center instead of a flat core, 
but the gravitational potential remains finite.
The potential produced by the NFW mass density profile may be referred to as the NFW potential.
Some other simulations also indicate similar density profiles with central cusps,
although the cusp slope may be somewhat different from that of equation (\ref{eq:nfw-dm}) (Moore et al. 1997; Fukushige and Makino 1997).

ASCA X-ray observations have shown noteworthy results concerning the potential structure 
in the central regions of nearby clusters.
In the Fornax cluster, 
Ikebe et al. (1996) found a hierarchical distribution of the total mass and dark matter.
In the Hydra-A, Abell 1795, and a few other clusters, 
Ikebe et al. (1997), Xu et al. (1998), and Xu (1998), respectively, 
revealed similar deviations from the King-type potential in their central regions.
Qualitatively these phenomena are reminiscent of the NFW potential, 
but the measured potential profiles appear to be more deviated from the King potential 
than can be explained by equation (\ref{eq:nfw-dm}).

It is therefore of great interest to examine the relation between the NFW potential and the hierarchical {\it ASCA} potential.
However, further assessing the reality of the NFW potential model is rather difficult.
This is because these clusters showing evidence of hierarchical potential structure all have cD galaxies at their centers, 
whereas the $N$-body simulations of NFW97 and others do not consider the significant baryonic component that must be associated with the cD galaxy.
In addition, 
most of these clusters show significant cool emission with a temperature of $\sim 1$ keV near the center, 
which introduces relatively large errors in the potential shape determinations via X-ray observations.

Accordingly, 
we consider that the center of ``non-cD clusters'' is the best place to compare the X-ray observations and the NFW prediction,
since these system are thought to have a relatively small fraction of baryonic matter and insignificant emission of cool component in the center.
We may then examine whether or not 
the high-quality X-ray data from {\it ROSAT} and {\it ASCA} of appropriate objects agree 
with predictions of the NFW potential.

We select the Abell 1060 cluster (A1060 for short),
based on our belief that it is the most suitable cluster for our purpose.
This cluster, with $z=0.011$,  is the nearest one after the Virgo, Fornax, and Centaurus clusters among X-ray luminous ones.
X-ray images of the cluster obtained with {\it Einstein} (Fitchett and Merritt 1988),  {\it ROSAT} and {\it ASCA}  have a good circular symmetry, 
justifying the assumption of the spherically symmetric ICM distribution to be employed in our analysis.
The symmetric image also ensures that there is no significant bulk motion of the ICM such as merger with substructure at the center of A1060.
There are two giant elliptical galaxies, NGC~3311 and NGC~3309, 
the former sometimes regarded as a cD galaxy.
However NGC~3311 is smaller in size and mass than the more typical cD galaxies in other nearby X-ray clusters.
In fact, 
the central excess X-ray luminosity above the $\beta$ model (Jones and Forman 1984), 
and the estimated cooling flow rate (Edge and Stewart 1991, Singh et al. 1988), 
are both very low in A1060 among nearby X-ray clusters.
Using the {\it ASCA} data, Tamura et al. (1996; hereafter T96) have shown that the ICM in central regions of A1060 is quite isothermal at $\sim 3.1$ keV with little evidence of cool emission component.
We therefore expect that the influence of the central galaxy (or galaxies) is rather small in A1060 than in other nearby cD clusters.

Under this motivation, we re-analyze in this paper the {\it ASCA} data of A1060, 
employing also the {\it ROSAT} data.
The paper is arranged in the following way.
In the next section we briefly describe the {\it ASCA} and {\it ROSAT} observations of A1060.
In \S3,
we evaluate spatially resolved spectra from the cluster
and confirm that the ICM is close to be isothermal within $20'$ of the center, 
in agreement with T96.
This result justifies the assumption of hydrostatic equilibrium of ICM and its single-phased treatment,
which are employed in the subsequent analysis.
In \S4, 
we investigate the radial brightness distribution assuming an isothermal ICM,
to estimate the ICM density profile and hence the potential profile of the cluster.
In \S 5,
allowing  a deviation from an isothermal condition,
we perform a combined analysis of the {\it ASCA} and {\it ROSAT} data, 
and constrain the potential profile.
In the last section, 
we discuss and summarize the obtained results.

Throughout this paper,
we assume the Hubble constant to be $H_0 = 70\ h_{70}$ km s$^{-1}$Mpc$^{-1}$,
and use the 90\% confidence level unless stated otherwise. 
The solar Fe/H ratio is taken to be $4.68 \times 10^{-5}$ by number. 
At $z = 0.011$, 1 arcmin corresponds to $13h_{70}^{-1}$ kpc.

\section{OBSERVATIONS}
{\it ASCA} observations of A1060 were performed on 1993 
June 28 and 29,
with the GIS (Gas Imaging Spectrometer; Ohashi et al. 1996, Makishima et al. 1996) in the PH normal mode 
and the SIS (Solid-State Imaging Spectrometer; Bruke et al. 1994; Yamashita et al. 1997) in the 4--CCD bright mode. 
After screening events using the standard data-selection criteria (T96), 
we have obtained the net exposure times of 36 ksec and 33 ksec for the GIS and SIS, respectively.
From these observations, some authors have already reported results of spatially sorted spectral analysis, 
including T96, Mushotzky et al. (1996), and Fukazawa et al. (1998).
In this paper we perform more detailed analysis of spatial variations of the X-ray spectrum, 
and derive constrains on the underlying potential.

\begin{figure}
\begin{center}
\leavevmode\psfig{figure=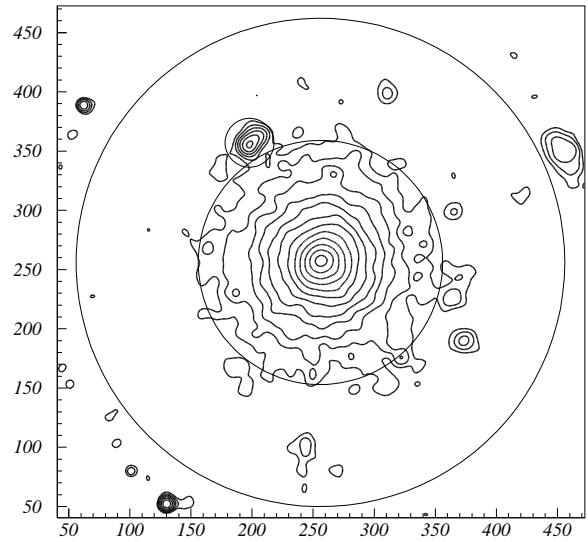,width=0.5\textwidth}
\figcaption{X-ray image of A1060 taken with the {\it ROSAT} PSPC in 0.5--2.0 keV. 
One pixel corresponds to $0'.25$.
The image has been smoothed with a Gaussian filter with $\sigma = 1'$, and 
corrected for the vignetting and partial shadows due to the detector support
ribs.
The contour levels are logarithmically spaced. 
The three circles show radii of $25'$ and $50'$ from the detector center, 
and the excluded region around HCG~48. 
North is up and east is to left.
\label{fig:pspc-image}}
\end{center}
\end{figure}

We also analyze the archival {\it ROSAT} PSPC data of A1060.
The observation was performed on 1992 January 1 with a net exposure time of 15.8\ ksec.
Imaging analysis using the data is described by Peres et al. (1998) together with those of other 55 clusters, 
and by Loewenstein and Mushoztky (1996).
Figure~\ref{fig:pspc-image} shows X-ray image of A1060
obtained with the PSPC in 0.5--2.0 keV.
The X-ray source seen at $29'$ north-east of A1060 
is the group of galaxies, HCG 48, which has a similar distance as the cluster does.
We exclude the data within $5'$ in radius of the X-ray center of HCG 48, 
when we investigate the surface brightness profile of A1060 with the PSPC.
The {\it ASCA} GIS image, presented in T96, is similar to the PSPC image.


\newpage
\section{SPECTRAL ANALYSIS}\label{sect:t-m-profiles}
In the present paper, we estimate the gravitating mass distribution of the cluster
under the assumptions of hydrostatic equilibrium and single-phase nature of the ICM.
Although we have selected A1060 as the target of our study believing that its ICM is nearly in an ideal condition, 
we must further examine the validity of these assumptions.
For this purpose, 
we investigate in this section the temperature structure of A1060 with the highest accuracy. 
Large scale temperature variations would suggest a cluster merger or substructure, 
and hence a deviation from the hydrostatic equilibrium of the ICM.
A cool  plasma component, 
ascribed to the ICM cooling or interstellar medium of central galaxies, 
may exist along with the hot ICM.
In such cases, the mass determination would become significantly less reliable.
In this section, we jointly use the X-ray spectra obtained from {\it ASCA} and {\it ROSAT}, 
to demonstrate,
with a higher reliability than was obtained by T96, 
that the ICM in A1060 is close to being isothermal.

\subsection{Large Scale Temperature Profile}\label{sect:tem-pro}
Utilizing the {\it ASCA} SIS and GIS data, 
T96 reported that the ICM in A1060 has spatially uniform temperature and metallicity 
within typical uncertainties of $\pm 10$\% and $\pm 30$\%, respectively.
However, they did not correct the {\it ASCA} data for the complex point-spread-function (PSF) of the X-ray telescope (XRT) onboard {\it ASCA}.
Accordingly, 
we re-analyze the {\it ASCA} data taking the PSF into account, 
and constrain the temperature profile with a higher reliability.
To evaluate a large scale temperature profile, 
in this subsection we use only the {\it ASCA} GIS data since it has a wider field of view than the {\it ASCA} SIS 
and a better spectral resolution than the {\it ROSAT} PSPC.
Since the X-ray brightness of A1060 is circularly symmetric, 
we accumulated X-ray spectra from five concentric ring regions ($0'-3'-6'-9'-12'-20'$) in the GIS detector plane centered on the X-ray centroid.
Due to extended tails of the PSF, 
each of these spectra contains photons scattered in from the other sky regions.
To correctly take into account this effect, 
we followed the method of Takahashi et al. (1995) and Ikebe et al (1997).
We divided the sky region into the corresponding five rings ($0'-3'-6'-9'-12'-20'$) around the cluster center.
For each energy bin, 
we calculated a 5 (sky annuli) $\times$ 5 (detector annuli) matrix, 
called image response matrix, 
which describes how photons from each sky region are distributed into the five detector regions.
In this calculation we assumed that the spectrum is uniform within each sky region.
Then, by specifying model spectra in the five sky regions (including their proper normalization), 
we can predict the five spectra on the detector plane, 
which can be fitted simultaneously to the actual five spectra.

In practice, we specified the brightness normalization in each sky region independently from each other.
For the spectral model, we employed the Raymond-Smith (Raymond and Smith 1977) plasma emission model modified by the photoelectric absorption.
The column density and metallicity were assumed to be constant over these regions, 
at the Galactic value of $6\times 10^{20}${\rm cm$^{-2}$} and 0.32 solar, respectively,
as obtained in T96.
Thus, the model involved 10 free parameters;
normalizations and temperatures of the five sky regions.
The background spectra were obtained from the blank-sky (containing no bright sources) database, 
by extracting events within the identical region from the same detector as the on-source data.
The background were added to the emission model.
When calculating the fit goodness,
we assigned 3\% and 10\% systematic errors to the spectral model and the background normalization, respectively.
The former is due to uncertainties in the energy response and PSF, 
while the latter represents errors in reproducing non-X-ray background and intrinsic fluctuation of the Cosmic X-ray background (Ishisaki 1996).

\begin{figure}
\begin{center}
\leavevmode\psfig{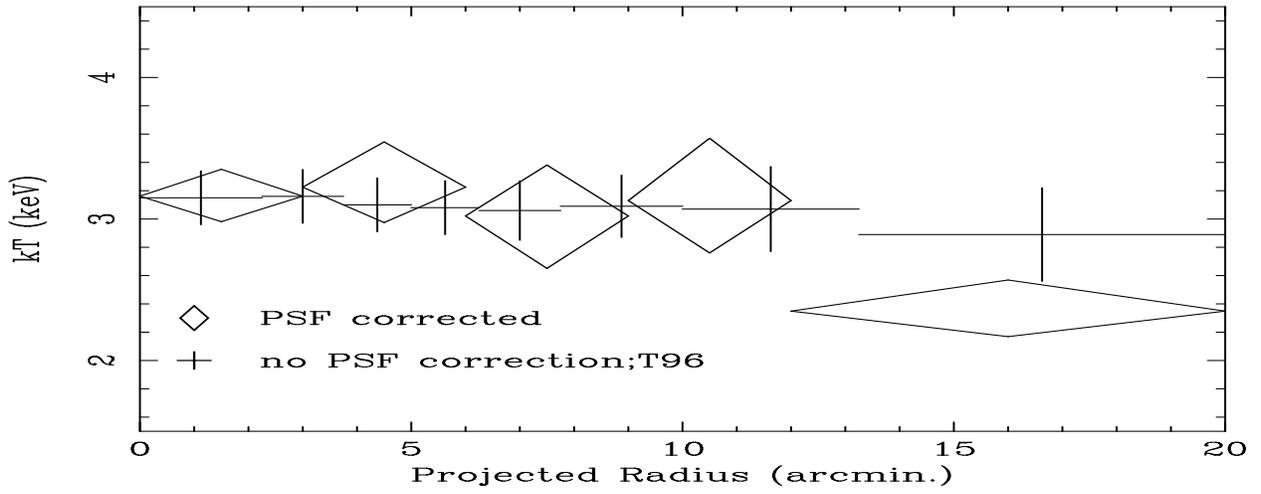}
\figcaption{ICM temperature vs. projected radius obtained from the PSF corrected method (diamonds). 
For comparison, the result from T96, in which no PSF correction was performed, 
is shown by crosses. 
\label{fig:2d-tr}}
\end{center}
\end{figure}

We first tested an isothermal model
in which all five temperatures are tied to be the same value, 
and found that it is roughly acceptable with $\chi^2 /\nu = 178/164$.
The obtained temperature is $3.07\pm 0.05$ keV, in agreement with the value obtained from the averaged spectra (T96).
We next let the five temperatures free.
The fit has been improved to $\chi^2 /\nu = 146/160$, 
and yielded roughly constant temperature within $12'$ in radius with a slight drop in the outermost annulus as presented in Figure~\ref{fig:2d-tr}.
According to an $F$-test, 
the fit improvement is significant at 84\% confidence level, indicating that the slight drop of temperature is marginally significant. 
As shown in Figure~\ref{fig:2d-tr}, 
effect of the PSF correction can be found only in the outermost region.
This can be explained as follows.
Due to the energy dependence of the PSF, 
higher energy photons are more heavily scattered off than softer ones.
Accordingly, the outer-region spectra become artificially harder than they are
and the possible outward temperature drop becomes less evident in T96.
Thus, 
correcting for the PSF, we have confirmed the inference made in T96, 
that the ICM in A1060 has an uniform temperature at least within $12'$ 
with a higher reliability.


\subsection{Possibility of the Central Cool Emission} \label{limit-cool}
In many cD clusters, 
significant cool emission is found in their central regions (e.g., Fabian et al. 1994).
In the case of A1060, we have shown that its ICM is close to an isothermal condition using the GIS data.
In addition, 
T96 gave a rather tight upper limit on the cool emission component at the cluster center using the GIS and SIS data.
To confirm the central cool component more tightly,
we further employ here the {\it ROSAT} PSPC data in addition to the {\it ASCA} data, 
because the PSPC has a higher sensitivity than the {\it ASCA} instruments to cooler emission components.
Therefore the joint fit using the three detectors, which covers the energy range of 0.3--10 keV, 
is an ideal method for our purpose.

We accumulated photons over the central region of radius of $5'$ (or $67 h_{70}^{-1}$ kpc), 
separately for the three detectors.
We chose this particular region because the cool emission is typically confined to central regions of $r<70 h_{70}^{-1}$ kpc in the Virgo (Matsumoto et al. 1996), Centaurus (Fukazawa et al. 1994), and AWM 7 clusters (Xu et al. 1997),
and because the angular size of $5'$ is large enough compared to the PSF of {\it ASCA}.
For the first step, 
we separately evaluated spectra obtained with the three detectors.
The GIS background was obtained in the same way as in \S~\ref{sect:tem-pro}.
That of the SIS was also obtained from the blank-sky observations.
On the other hand, 
the PSPC background was accumulated from the region of radius $36'-46'$ in the field of view.
We estimate the cluster emission in this background region to be $\sim 1$\% of that in the central region (R$<5'$) based on extrapolation of the radial brightness profile.
Therefore the cluster contribution in the background spectrum is negligible.
These background spectra were subtracted from the data before fitting to the model.
The temperature, column density and metallicity were all allowed to be free, 
and different from instrument to instrument.
In Table~\ref{tbl:r5spectra},
we show the result of fitting with a single temperature Raymond-Smith model.
We consider that the relatively larger column density obtained with the SIS
is due to the response uncertainty below 1 keV of the detector;
the fitting results with the SIS in other objects show slightly large column densities 
than those with the GIS\footnote{See e.g., a calibration status memo in http://heasarc.gsfc.nasa.gov/docs/asca/ahp\_proc\_analysis.html, 
which is maintained at NASA/GSFC.}.
Although the GIS and SIS temperature determinations are in a good agreement,
the PSPC temperature is somewhat lower, 
and disagree with the {\it ASCA} values at the 90\% level.

The lower temperature obtained with the PSPC suggest the presence of plasmas,
cooler than the global ICM, at the central region.
To examine this possibility, 
we fitted the PSPC spectrum by adding another plasma component to the model (i.e., two temperature model).
If we leave the two temperatures free, 
the two components are coupled too strongly with each other.
Therefore we fixed the temperature of hot component at 3.1 keV which is a global temperature of the cluster determined with the {\it ASCA}.
The metallicity of both components were also fixed at the global value of 0.3 solar derived from the {\it ASCA} spectra.
This two temperature model gave slightly better fit ($\chi ^2/\nu = 39/37$) to the data.
The temperature of the cool component is fount to be 1.1 (0.9--1.5) keV.
However, 
the fit improvement is significant only at 40\% confidence level based on the $F-$test.
Adding a cooling flow model instead of the cool plasma model did not make the fit better significantly($\chi ^2/\nu = 41/36$).

Alternatively, the lower PSPC temperature may be an indication of temperature decrease toward the center on small scales.
To examine this possibility,
we evaluated the PSPC spectrum within a smaller radius of $2'.25$.
A single temperature fit gave a temperature of 2.2 (1.8--2.7) keV, 
which is similar to that obtained previously from the region of $5'$ radius.
Two temperature model to this spectrum did not 
give significant improvement of the goodness of the fit ($\chi ^2/\nu = 119/133$ vs. $\chi ^2/\nu = 119/131$).
Thus, the PSPC spectra show no significant temperature decrease on a few arcmin scale.

\begin{figure}
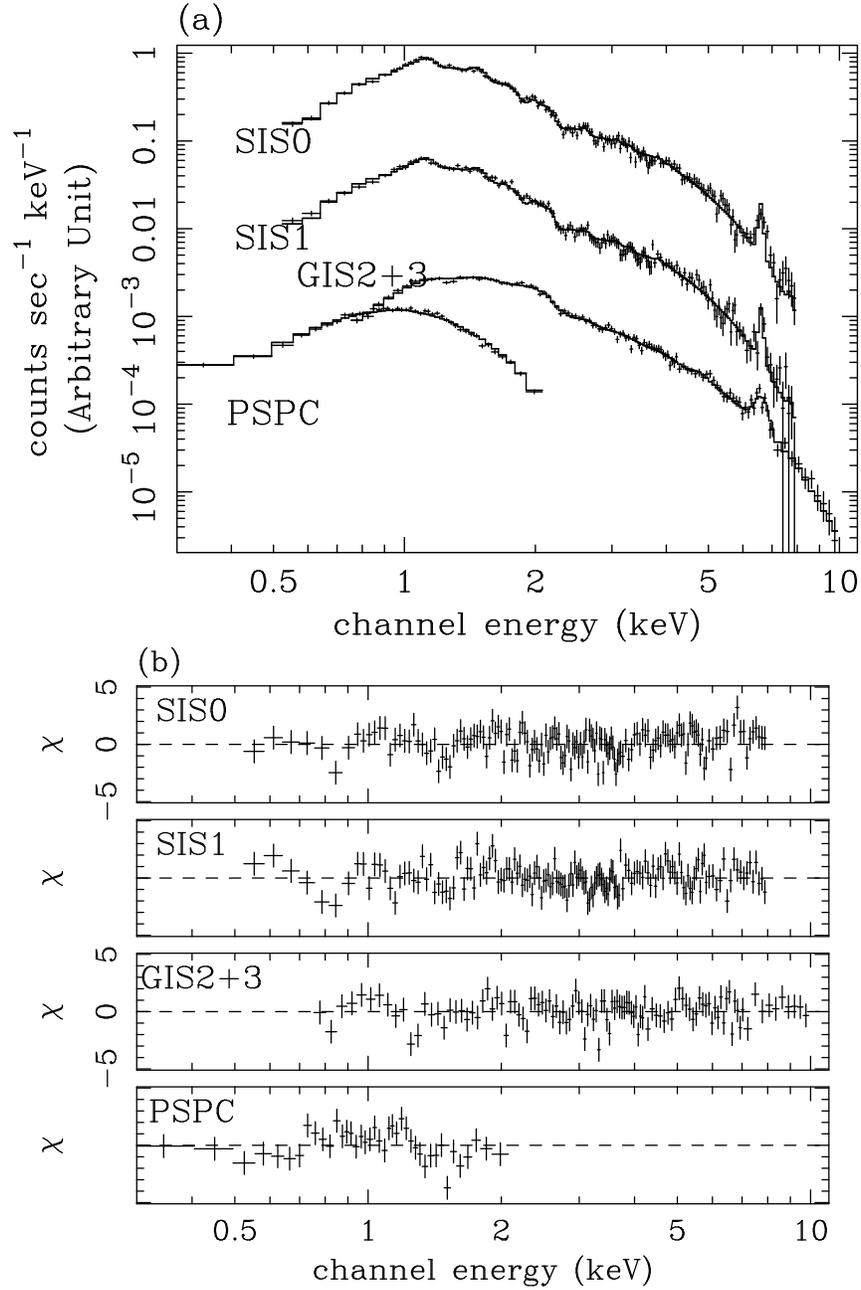

\begin{center}
\leavevmode\psfig{figure=fig3a.ps,angle=-90,height=0.4\textheight}
\leavevmode\psfig{figure=fig3b.ps,angle=-90,height=0.4\textheight}
\figcaption{Results of the joint fit to the {\it ASCA} and {\it ROSAT} spectra of the central $5'$ of A1060. 
(a) The spectra from the SIS, GIS, and {\it ROSAT} PSPC are shown together with the best-fit model.
The four spectra are vertically offset for clarity.
(b) The fit residual for each detector, 
SIS-S0, SIS-S1, the GIS, and the PSPC, are shown in unit of standard deviation.
\label{fig:joint-sp-r5}}
\end{center}
\end{figure}

For the next step, 
we fitted jointly the three spectra (SIS+GIS+PSPC) with a single temperature model
to further examine the isothermality near the cluster center.
The model temperature and metallicity were constrained to be common among the three detectors.
On the other hand, the normalization and column density 
were allowed to take independent values, 
to take into account the slight mismatch in absolute photometric calibration 
and systematic uncertainties in low energy detector responses, 
respectively.
This has given an roughly acceptable fit with $\chi^2/\nu = 556/459$, yielding 
a temperature of $3.3\pm0.1$ keV, 
and a metallicity of $0.29\pm0.03$ solar (Table~\ref{tbl:r5spectra} and Figure~\ref{fig:joint-sp-r5}).
This implies that the emission within a projected radius of $5'$ is approximately isothermal.

In a search for the possible cool emission,
we added another plasma component
to jointly fit the spectra from the three detectors.
The metallicity of both components were assumed to be the same,
while the two temperatures  were left free.
The fit was in fact improved slightly from 556/459 to 537/457
by introducing the cooler component, 
as compared to the single temperature fit.
Therefore, a cool component may in fact be present, 
as suggested by the lower PSPC temperature.
However, the improvement is significant only at 32\% confidence level based on the $F$-test.
We therefore quote a conservative 90\% upper limit on the cool-component emission measure 
as 4\% of that of the hot component,
when the cool component has a temperature of 1 keV.
This is in a good agreement with T96.

Based on these results, 
we conclude that the {\it ASCA} and {\it ROSAT} PSPC spectra do not require strongly an additional cool component.
Consequently, we assume a single-phased ICM in the subsequent analysis.

\placetable{tbl:r5spectra}

\section{RADIAL BRIGHTNESS DISTRIBUTIONS}\label{sect:radial}
In this section we evaluate the X-ray surface brightness distribution of A1060 and derive the ICM density profile.
In the previous section,
we found that the X-ray emission is dominated by an isothermal component of $\sim 3$\ keV.
Therefore, the brightness directly relates to the ICM density and hence to the gravitational profile of the cluster.
We separately analyze the X-ray images obtained with the {\it ASCA} GIS and the {\it ROSAT} PSPC, 
but we do not use the {\it ASCA} SIS data because of its limited field of view.
Since the GIS and the PSPC have different energy bands,
a comparison of the brightness distributions from the two instruments provides
another estimation of the temperature distribution.
If the ICM is actually isothermal, 
the two brightness should be similar to each other.

\subsection{Model Fittings to the GIS Data}
We derived an azimuthally averaged count-rate profile from the two GIS detectors (S2 and S3), 
centered on the X-ray peak, in the 0.7--10 keV band.
We quantify the profile in ``forward'' way;
we start from a model surface brightness distribution,
apply the XRT vignetting to it,
and further convolve it with the XRT$+$GIS point spread function.
The background, 
obtained in the same way as in \S~\ref{sect:tem-pro}, 
is added to the model.
The derived model prediction is compared with the observed count-rate profile.
The image response matrices were utilized again to take the spatial response into account in reproducing the model-predicted count profiles.
In the present case, 
we employed matrices of 20 (sky annuli) $\times$ 20 (detector annuli) in size.
According to the results obtained in T96 and in \S\ref{sect:t-m-profiles} of this paper,
we assumed a model X-ray spectrum with temperature of 3.1 keV and metallicity of 0.32 solar, 
absorbed with the Galactic column density of $6\times 10^{20}$ cm$^{-2}$.
We assigned 3\% and 10\% systematic errors to the count-rate models and background estimation, 
respectively, as in \S\ref{sect:t-m-profiles}.

As the model ICM density profiles, 
we considered the following two cases.
One is the case where an isothermal ICM is gravitationally confined in a King-type potential (e.g., Sarazin 1988).
The model density in this case becomes the usual $\beta$ model, as
\begin{eqnarray}
\rho^{\rm beta}_{\rm icm}(r)	 \propto \left[1 + \left(\frac{r}{r_{\rm c}}\right)^2\right]^{-1.5 \beta}, \label{eq:beta-gas2}
\end{eqnarray}
where $r_{\rm c}$ is the core radius and the parameter $\beta$ usualy takes a value $\sim 0.7$.
The other is the case in which an isothermal ICM is confined in the NFW potential corresponding to equation (\ref{eq:nfw-dm}).
As calculated by Makino, Sasaki, and Suto (1998), 
the ICM in this case exhibits a radial density profile as 
\begin{eqnarray}
\rho^{\rm NFW}_{\rm icm} (r)  \propto & \left(1+\frac{r}{r_{\rm s}}\right)^{\frac{B r_{\rm s}}{r}} \label{eq:nfw},
\end{eqnarray}
where $B$ is a parameter related to the ICM temperature and $r_{\rm s}$ is a scale length.
Although this model form appears quite different from that of the $\beta$ model, 
in reality it can fairly closely mimic the $\beta$ model profile, 
especially outside the core region (Makino et al. 1998).
We call this density profile the NFW ICM model.

\begin{figure}
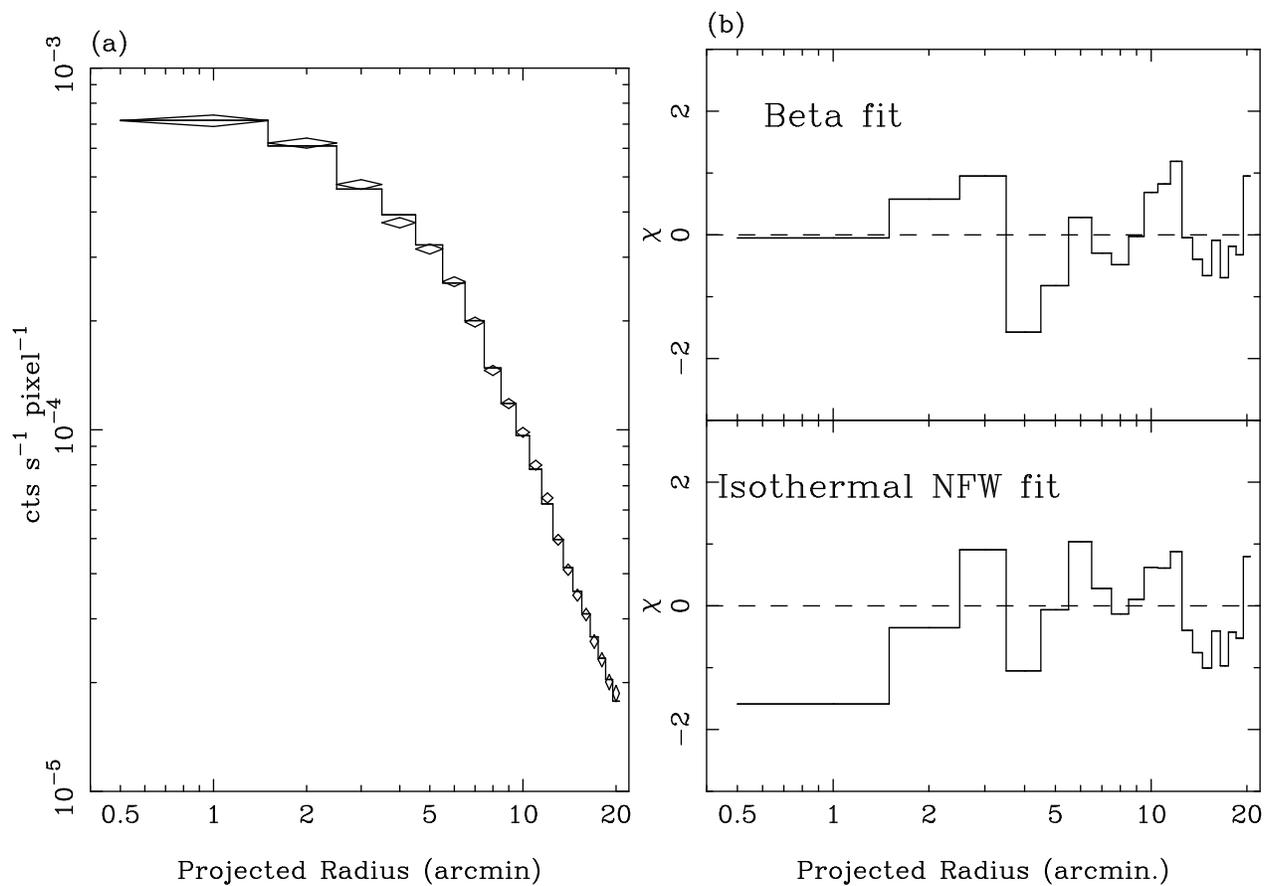

\centerline{\hbox{
\psfig{figure=fig4a.ps,width=0.5\textwidth}
\psfig{figure=fig4b.ps,width=0.5\textwidth}
}}
\figcaption{The $\beta $ and NFW ICM model fits to the 0.7--10 keV GIS count-rate profile of A1060. 
(a) The data and the best-fit $\beta$ model represented by the diamonds and histograms respectively.
Note that the estimated background was added to the model.
(b) The residual from the $\beta$ fit (top) and from the NFW ICM model (bottom) in unit of standard deviation.
\label{fig:gis-rp}}
\end{figure}

As shown in Table~\ref{tbl:rpfit} and Figure~\ref{fig:gis-rp},
both the $\beta$ model and the NFW ICM model have given acceptable fits to the GIS count-rate profiles.
The obtained core radius and $\beta$ are consistent with those obtained with {\it Einstein} (Jones and Forman 1984).
Although Jones and Forman (1984) found a central excess above the $\beta$ model in the {\it Einstein} IPC data of A1060, 
the GIS data do no indicate significant excess in the central region.
This is probably because the IPC has a better spatial resolution (a half power diameter of $\sim 1'.2$) than {\it ASCA}.


\subsection{Model Fittings to the PSPC Data} \label{sect:pspc-density}
To further investigate the ICM density profile, 
we utilize the {\it ROSAT} PSPC data.
The PSPC has a better spatial resolution of $\sim 25''$ in half power diameter, compared to {\it ASCA}.
Accordingly, we expect to obtain stronger constraints on the ICM density profile.

We followed a standard method (Zimmermann et al. 1997) to convert the archival PSPC data of the cluster to a surface brightness profile in the 0.5--2.0 keV range.
We derived a radial count-rate profile by accumulating photons within a number of ring-shaped regions in $30''$ bins.
Dividing this count rate profile by an exposure map, 
we obtained an azimuthally averaged surface brightness profile as a function of projected radius.
The cluster emission is detected at least up to radii of $\sim 30'$ with signal to background ratios larger than 0.2.
We did not consider the finite width of the PSF of the PSPC, 
since its scale is similar to or less than the binning of the count-rate profile.
Therefore, 
we cannot discuss structures smaller than that of the PSF.
In other words, 
we assume the image response matrix to be diagonal.
Since estimation of background is rather difficult with the PSPC,
background was added to the fitting model as a flat component, with a free normalization and a 5\% systematic uncertainty.

\begin{figure}
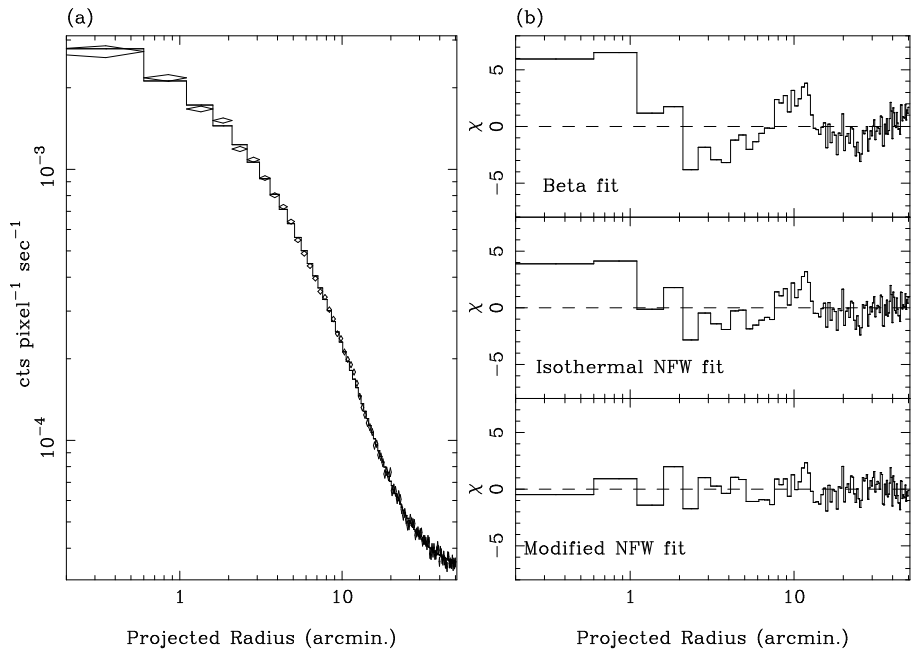

\begin{center}
\leavevmode\psfig{figure=fig5a.ps,height=0.4\textheight}
\leavevmode\psfig{figure=fig5b.ps,height=0.4\textheight}
\figcaption{
(a) The 0.5--2.0 keV PSPC radial profile along with the modified NFW ICM best-fit model.
(b) The residuals from 
the $\beta$ model, isothermal NFW model, and the modified NFW model, respectively, 
from top to bottom in unit of standard deviation.
\label{fig:nb-fit2}}
\end{center}
\end{figure}

We fitted the obtained PSPC profile first with the $\beta$ model.
As shown in Table~\ref{tbl:rpfit} and illustrated in Figure~\ref{fig:nb-fit2}, 
the $\beta$ model has been rejected with a significant confidence.
This is because of a significant data excess above the model within a radius of $1'$.
Then we fitted the data with the NFW ICM model.
This model described the data better than the $\beta$ model, 
although the fit is still formally unacceptable,
as shown in Table~\ref{tbl:rpfit} and illustrated in Figure~\ref{fig:nb-fit2}.

\subsection{A Modified ICM Model} 
The observed PSPC brightness still exhibits excess in the central region above the NFW ICM model.
To reproduce this excess,
we modified the NFW ICM model function by introducing another parameter $n$ as
\begin{eqnarray}
\rho ^{\rm NFW'}_{\rm icm} (r)& \propto & \left[1+\left(\frac{r}{r_{\rm s}}\right)^n\right]^{\frac{B r_{\rm s}}{r}}  \label{eq:md-nfw-gas2}.
\end{eqnarray}
The case $n=1$ corresponds to the NFW ICM model.
The smaller the index $n$, the steeper the ICM density of equation(\ref{eq:md-nfw-gas2}) becomes within $\sim r_{\rm s}$  of the center.
Although this function form is not quite simple, 
it is a natural generalization of the NFW ICM model and $\beta$ model on the outer region of $r/r_{\rm s} > 1$ on condition of $n > 0.95$.
We hereafter call equation~(\ref{eq:md-nfw-gas2}) modified NFW ICM model.
We fitted this model to the PSPC brightness profile and obtained an acceptable fit with $n=0.97$ (Table~\ref{tbl:rpfit} and Figure~\ref{fig:nb-fit2}).

These results imply that 
the ICM of A1060 is more concentrated towards the cluster center than is predicted by the $\beta$ model.
The fitting result with the NFW ICM model suggests that the ICM is even more concentrated towards the center than is predicted by the NFW potential and the isothermal assumption.
Consequently, 
the gravitational potential is deeper than the NFW model [eq.(\ref{eq:nfw-dm})],
or the ICM temperature distribution deviates from the isothermal condition.
More quantitatively, 
an isothermal ICM takes the form of eq.(\ref{eq:md-nfw-gas2}) when it is confined with a gravitational potential of which the total mass density is given as
\begin{eqnarray}
\rho^{\rm NFW'}_{\rm tot} & \propto & x^{2n-3}(1+x^n)^{-2}\left[1+\left(\frac{1-n}{n}\right)\frac{1+x^n}{x^n}\right] \label{eq:md-nfw-tot}
\end{eqnarray}
with 
$x\equiv \frac{r}{r_{\rm s}}$.
In Figure~\ref{fig:nfw'-mass} we present this total mass density profile with $n=0.97$, 
as obtained above, and its integral form.
Thus, $\rho^{\rm NFW'}_{\rm tot}$ with $n=0.97$ is approximated as $r^{-1.5}$, 
compared to the original NFW profile which scales as $r^{-1.0}$.

We also note that the best-fit $\beta$ model parameters to the PSPC profile
and those to the GIS (Table~\ref{tbl:rpfit}) are similar,
even though the former is unacceptable.
This suggests that the central excess brightness found with the PSPC is relatively independent of X-ray energy.
This is because if, e.g., the central excess is due to a central temperature drop, 
we would have obtained a smaller core radius and a flatter $\beta$ with the PSPC than with the GIS.
We shall further address these issues in the next section.

\placetable{tbl:rpfit}


\newpage
\section{COMBINED SPECTRAL AND SPATIAL ANALYSIS:MASS CALCULATION}
\subsection{Motivations and Method}
In this section, 
we estimate the total gravitating
mass distribution allowing a deviation from isothermality of the ICM within tolerance of the {\it ASCA} plus {\it ROSAT} spectroscopy.
To do this, 
we consider the three dimensional temperature 
and density profiles of the ICM together, by combining the spectral and spatial analyses.
The temperature profile was actually constrained in \S\ref{sect:t-m-profiles},
but it was projections onto two dimensions.
Furthermore, the derived ICM density profile might change
if we properly take into account small variations
in the X-ray spectrum across the cluster,
which was neglected in \S\ref{sect:radial}.

We hereafter utilize the GIS spectra 
and the PSPC brightness profile {\em simultaneously}.
This is because the {\it ASCA} has superior spectroscopic capabilities
while the {\it ROSAT} has a better angular resolution.
There has so far been no such attempts to our knowledge, except for Ikebe et al. (1999),
even though many investigators use the {\it ASCA} and {\it ROSAT} data jointly.
We do not attempt to analyze the SIS data for the same reason as mentioned in \S\ref{sect:tem-pro} and \S\ref{sect:radial}.

In order to find a cluster model that can simultaneously reproduce
the {\it ASCA} spectra and the {\it ROSAT} radial brightness profile,
we have used a new analysis scheme.
This is a kind of variation of the method of Hughes (1989), 
who examined the mass of the Coma cluster using 
spatially averaged spectra from {\it Tenma} and {\it EXOSAT}
and X-ray imaging data from the {\it Einstein} IPC.
Markevitch and Vikhlinin (1997) adopted a similar method to derive the total mass in Abell 2256.

The procedure consists of the following steps.

\begin{itemize}
\item[1.]
We assume the total mass density $\rho_{\rm tot}(r)$,
and the ICM mass density $\rho_{\rm icm}(r)$,
both in some spherically-symmetric analytic forms 
as a function of the three-dimensional radius $r$.
Assuming $\rho_{\rm tot}(r)$ is equivalent to 
assuming the integrated total mass profile,
or the gravitational potential.
For simplicity's sake, we assume a single component total mass made of dark matter,
instead of multi-component mass modeling considering baryonic effects. 
This is reasonable because previous analysis showed that the dark matter dominates the total mass of this cluster within the observed region (e.g., Loewenstein and Mushotzky 1996).

\item[2.]
The condition of hydrostatic equilibrium relates
the total gravitating mass $M_{\rm tot}(<r)$ within $r$, 
$\rho _{\rm icm}(r)$, 
and pressure $P = \frac{\rho_{\rm icm}(r)kT(r)}{\mu m_p}$ where $\mu$ and $m_p$ are the mean molecular weight and proton mass, respectively, as 
\begin{eqnarray} 
\frac{dP}{dr} & = & - \frac{GM_{\rm tot}\rho _{\rm icm}}{r^2}. \label{eq:hydro}
\end{eqnarray}
We can integrate this equation outward from $r=0$ 
for a given set of temperature at $r=0$, 
$\rho_{\rm tot}(r)$, and $\rho_{\rm icm}(r)$ to obtain $P(r)$ and $T(r) \propto \frac{P(r)}{\rho_{ICM(r)}}$.
However, the solution of $T(r)$ is very sensitive to the choice of $T(0)$ or the normalization of the total mass (Huges 1989; Loewenstein 1994).
Therefore, following Loewenstein (1994), we rewrite equation(\ref{eq:hydro}) as 
\begin{eqnarray} \label{eq:mtr} 
\frac{dP}{dz}  = - GM_{\rm tot}\rho _{\rm icm} &{\rm with} & z \equiv \frac{1}{r}.
\end{eqnarray}
Integrating this equation from $z=0\ (r\rightarrow \infty)$ with a boundary condition of $P(z=0) \equiv P_{\infty}$, 
we can compute $T(r)$ numerically without specifying $T(0)$.
The non-zero pressure at infinity ($P_{\infty} >0$) would imply that the temperature goes to infinity at large radii.
We consider such solutions unphysical and assume $P_{\infty}=0$.
Note that $T(r)$ should take a particular form, 
in order for the assumed ICM to be in a hydrostatic equilibrium 
in the assumed potential via equation(\ref{eq:hydro}),
since $\rho_{\rm tot}(r)$ and $\rho_{\rm icm}(r)$ are independently specified in advance. 

\item[3.]
According to the specification of $\rho_{\rm icm}(r)$ and $T(r)$, 
and utilizing the Raymond-Smith emission code,
we analytically model the X-ray emissivity 
as a function of $r$ and energy.
We assume a constant metallicity of 0.32 solar 
and the Galactic column density over the entire cluster.
We transform the model emissivities into 
a set of expected spectra obtained with the GIS, 
using the image response matrices
which take into account the projection effects and the XRT+GIS response.
The model cluster is also converted to the simulated PSPC brightness in the 0.5--2.0 keV range.

\item[4.]
These model predictions are simultaneously fitted to the 
GIS ring-sorted spectra (5 regions; $0'-3'-6'-9'-12'-20'$) and the PSPC radial surface brightness profile up to the projected radius of $50'$ (670$h_{70}^{-1}$ kpc).
The fitting to the five GIS spectra takes into account not only their spectral shapes, 
but also their relative normalizations.
The model goodness is calculated through the $\chi^2$-evaluation. 
Systematic errors were assigned as in the previous fittings.
If necessary, the initial models ($\rho_{\rm tot}$ and $\rho_{\rm icm}$)
are corrected in order to improve $\chi^2$.
\end{itemize}

Instead of the above approach, 
we could first assume the temperature profile in some analytic forms.
In fact, Henriksen and Mushotzky (1986) employed a form of
$T(r) \propto \rho ^{\gamma}$,
and David et al. (1995) assumed $T (r) \propto r^{\alpha}$,
with $\gamma$ and $\alpha$ both being parameters.
However, such a parameterization of temperature along with a particular ICM density model severely constrains 
the range of mass profile.
Since our primary goal is constraining the total mass, 
or its density, rather than examining the temperature profiles, 
we first assume mass distributions as described above, 
and then determine temperature profile in such a way that the implied spectra are consistent with the observed data.

\subsection{Results}
\subsubsection{The NFW Potential Model}
Using the method described above, 
we examine the NFW potential model [eq.(\ref{eq:nfw-dm})] with a central density cusp.
For the ICM density, 
we assume the modified NFW ICM model [eq.(\ref{eq:md-nfw-gas2})] which has been found in the previous section to be the best representation of the PSPC radial profile.
We denote the scale radius of the ICM density as $a$ and allow $a$ to take different value from that of the total mass density $r_{\rm s}$.
The case $n=1$ and $a = r_{\rm s}$ corresponds to an isothermal ICM distribution.

The best-fit model has been obtained with $r_{\rm s}= 14'.6$ ($190 h_{70}^{-1} $kpc), $a=35'.6, B=10.2$, and $n=0.97$.
This model is acceptable with $\chi^2/\nu = 242/262$.
As shown in Figure~\ref{fig:gn-10},
 the solution reproduce both the {\it ASCA} spectra and the {\it ROSAT} radial brightness profile well.
The value of $n$ agree with that obtained in \S\ref{sect:radial} assuming isothermality.
The acceptable (99\% confidence) range  of the scale radius of the total mass is $12' <r_{\rm s} <19'$, 
which also overlaps considerably with those derived in \S~4 with the GIS or PSPC.
The total mass density and temperature profiles for a range of acceptable models are plotted by solid lines in Figure~\ref{fig:dens-tem}.
These results are reliable only within $\sim 20'$, 
because the GIS data covers up to $20'$.
The implied temperature exhibits a moderate drop within $\sim 3'$, 
because the best-fit ICM profile with $n=0.97$ increase towards the center more steeply than an isothermal ICM ($n=1$) in the NFW potential.


\subsubsection{The Power-law Potential Model}
\begin{figure}
\leavevmode\psfig{figure=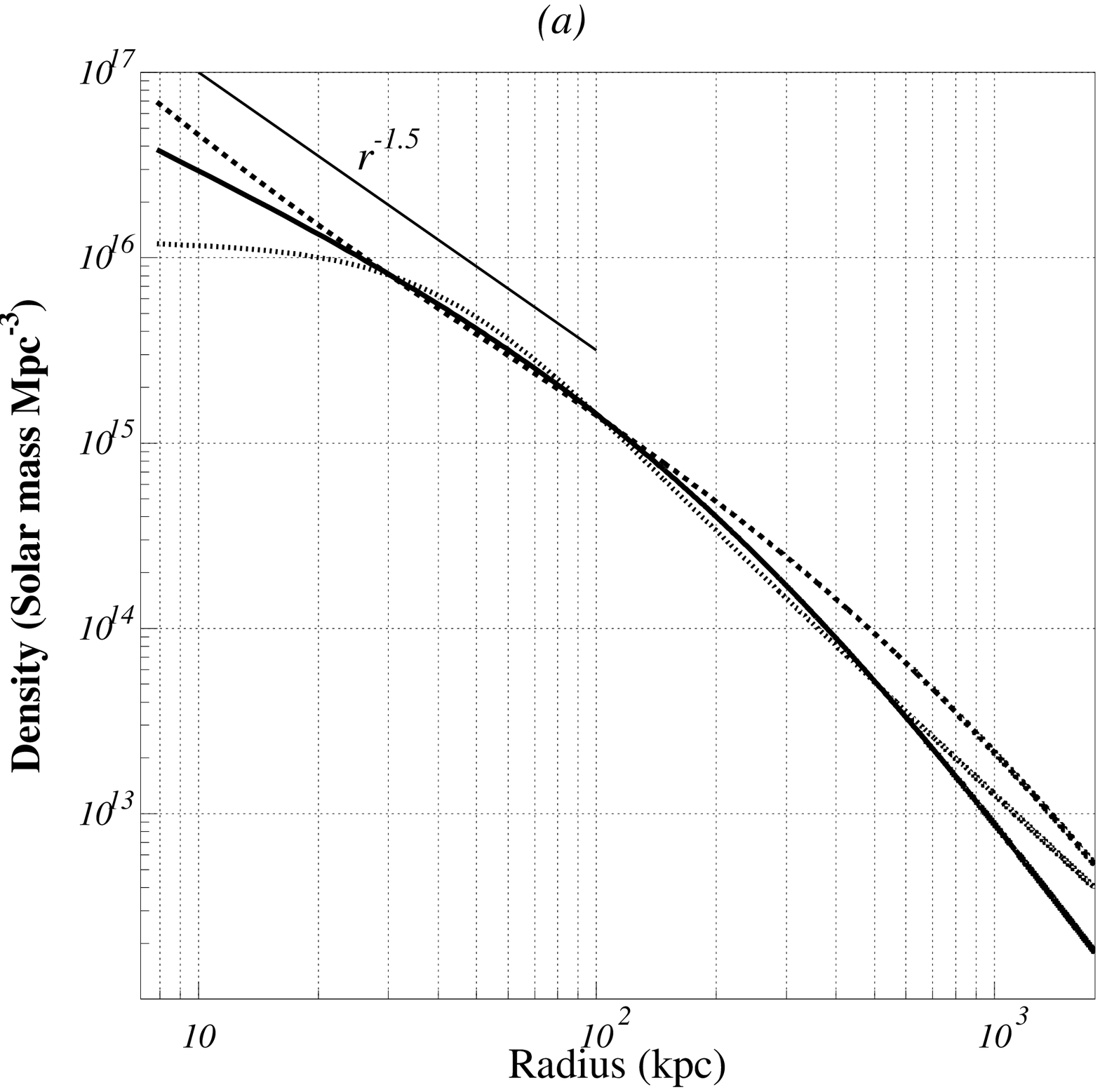,height=0.3\textheight}
\leavevmode\psfig{figure=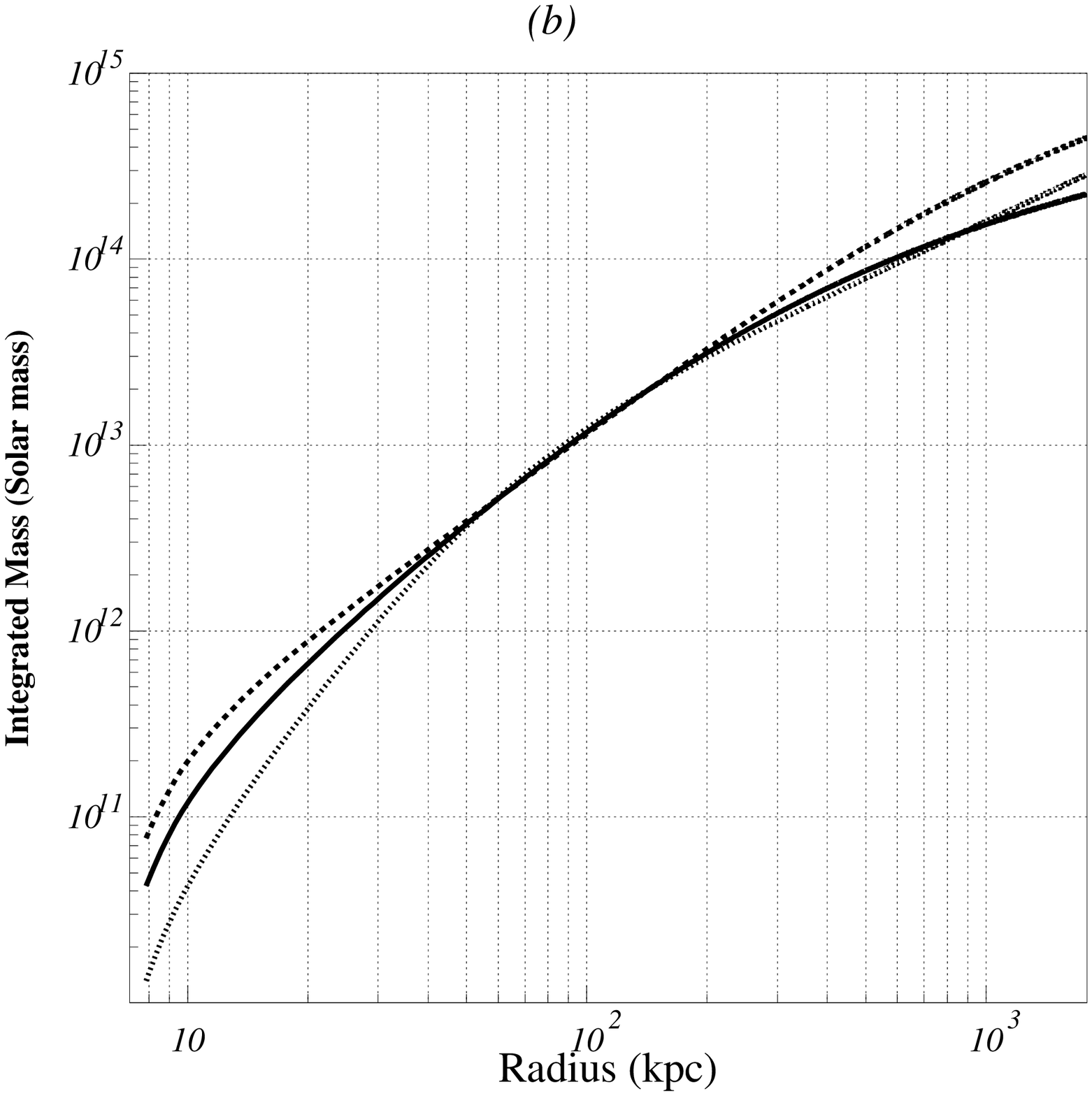,height=0.3\textheight}
\figcaption{
The total mass density (a) and its integrated representaion (b),
against the 3-dimensional radius,
derived from the surface brightness fitting to the PSPC radial profile assuming isothermality of the ICM.
The solid lines, dashed lines, and dotted lines show models based on the NFW ICM, modified NFW ICM, 
and $\beta$ model, respectively.
Only the second one gave an acceptable fits to the data.
A slope proportional to $r^{-1.5}$ is shown by a solid line.
We assumed $h_{70}=1$.
\label{fig:nfw'-mass}}
\end{figure}

\begin{figure}
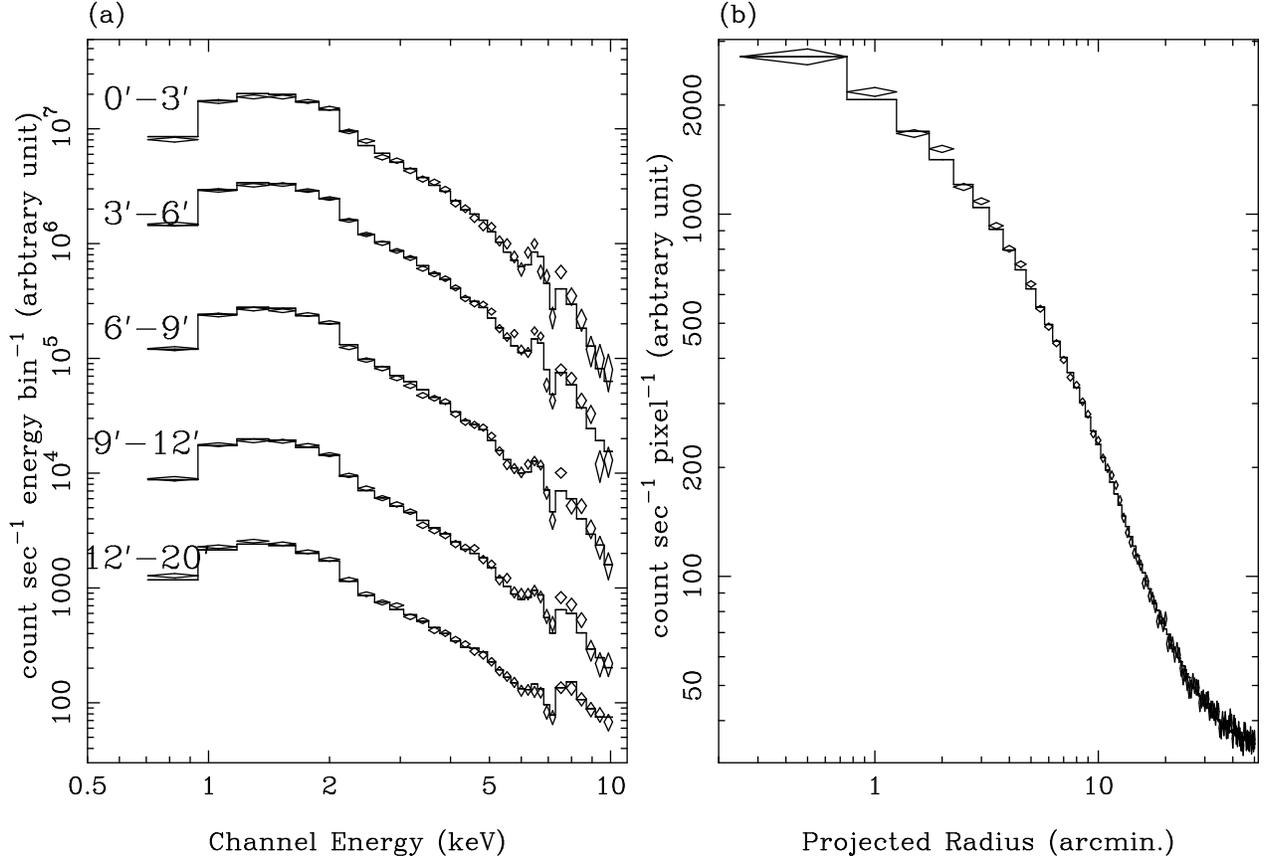

\centerline{\hbox{
\psfig{figure=fig7a.ps,width=0.5\textwidth}
\psfig{figure=fig7b.ps,width=0.5\textwidth}
}}
\figcaption{The best-fit solution in terms of the NFW total density model with $r_{\rm s} = 14'.6$ and the modified NFW ICM model with $n=0.97$.
(a) The GIS spectra accumulated from $0'-3'-6'-9'-12'-20'$ are shown by diamonds,
together with the model by histograms.
The five spectra are vertically offset for clarity.
(b) The PSPC radial profile (diamonds) and the model (histograms).
\label{fig:gn-10}}
\end{figure}

In \S\ref{sect:pspc-density}, 
we found the PSPC radial brightness to be even more
concentrated towards the center than is predicted by the NFW potential.
Therefore, the underlying potential is inferred to exhibit a stronger central drop than the simple NFW model.
Hence, we examine a total mass distribution with a steeper slope near the center than the NFW one.
We assume a total mass density model of the form
\begin{eqnarray}
\rho _{\rm tot}^{\rm Power} & \propto  & \left(\frac{r}{r_{\rm s}}\right)^{-\eta} \left(1+\frac{r}{r_{\rm s}}\right)^{\eta - 3} \label{eq:nfw-dm2},
\end{eqnarray}
where $\eta$ is a free parameter, 
and the case of $\eta = 1$ corresponds to the NFW model of equation(\ref{eq:nfw-dm}).
Since the density slope $\eta$ and scale radius $r_{\rm s}$ were difficult to be determined separately,
we fixed $r_{\rm s}$ to be $100' \sim 1400 h_{70}^{-1}$ kpc.
In this case,
the total mass density is close to a power-law form of 
$\rho_{\rm tot} \propto r^{-\eta}$
in the central region of the cluster ($r<20'$), where we are interested in.
We refer to this the power-law density model.
This profile approximates the total mass density of eq.(\ref{eq:md-nfw-tot}), 
derived in \S~\ref{sect:pspc-density}
based on the PSPC radial brightness profile and the isothermal assumption (Figure~\ref{fig:nfw'-mass}).
The formula implies that the total mass diverges in proportion to $\log(r)$ as the NFW model.
When an isothermal ICM is confined within this power-law density potential, 
the ICM density profile becomes very close (though not exactly identical) to 
the modified NFW ICM profile of equation(\ref{eq:md-nfw-gas2});
$n=0.97$ corresponds to $\eta \sim 1.5$.
Therefore, we express the ICM density profile by equation(\ref{eq:md-nfw-gas2}), 
to be combined with the total density profile of equation(\ref{eq:nfw-dm2}).

\begin{figure}
\centerline{\hbox{
\psfig{file=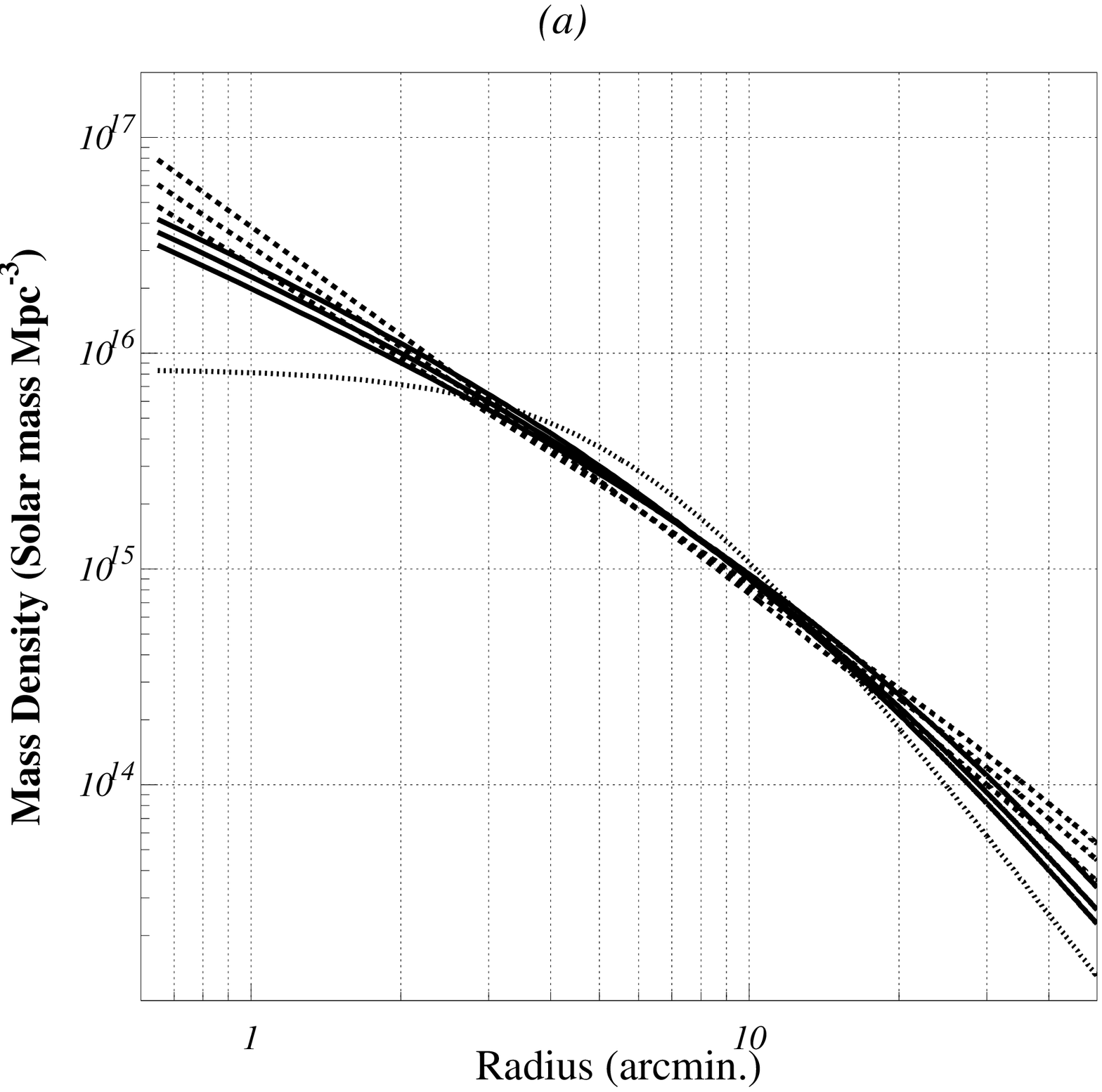,width=0.5\textwidth}
\psfig{file=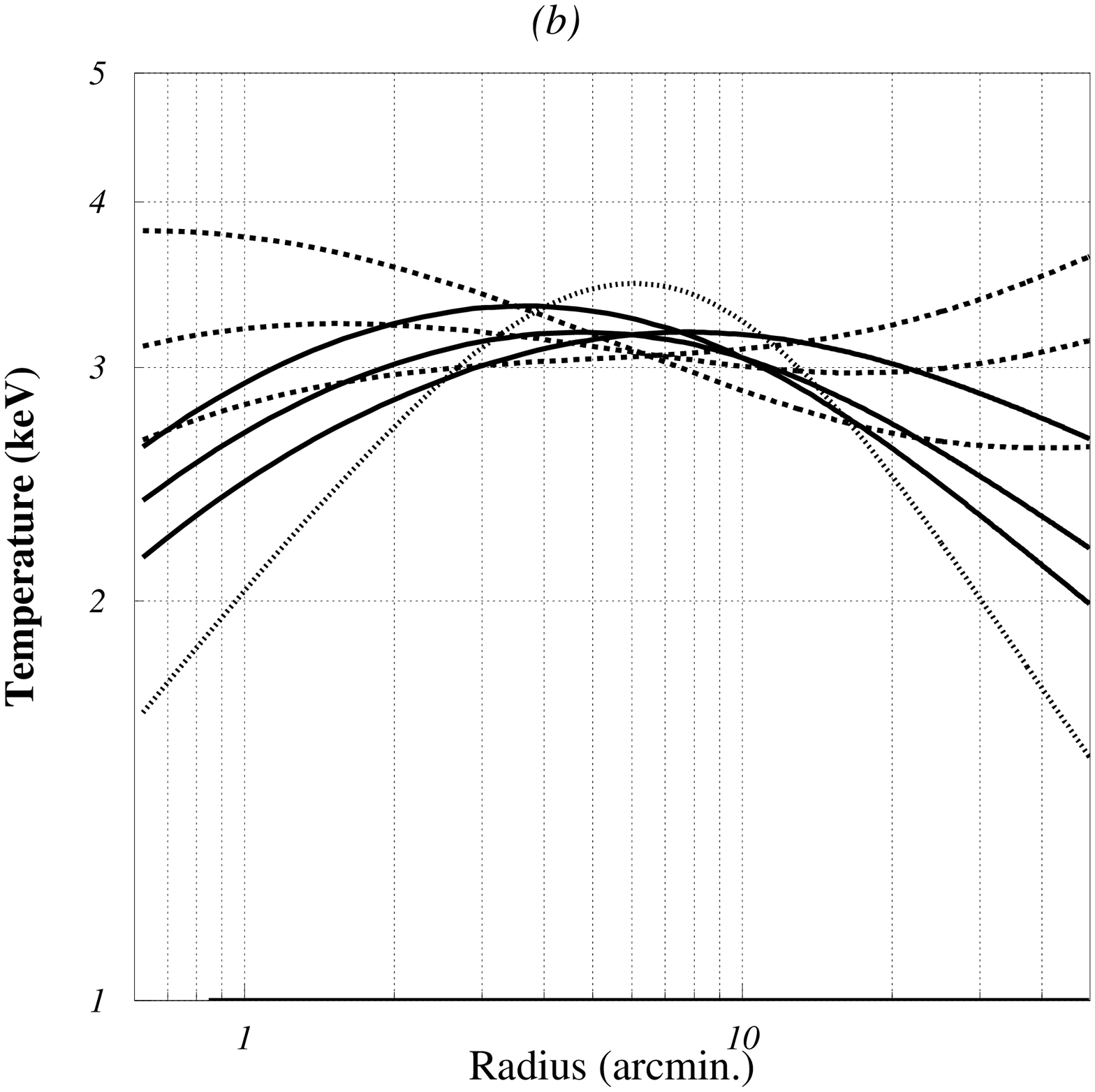,width=0.5\textwidth}
}}
\figcaption{The bset-fit and acceptable models from three potential profiles
derived through a simultaneous fitting to the GIS and PSPC data.
Panels (a) and (b) show the total mass density profiles and 
corresponding temperature profiles, respectively.
The solid, dashed, and dotted lines show 
the NFW, power-law density, and King-type potential models, respectively.
Note that the King-type model does not give acceptable fits to the data.
\label{fig:dens-tem}}
\end{figure}

The best-fit model has been obtained with $\eta=1.53, a=35'.1, B=10.1$, and $n=0.97$ (dashed-lines in Figure \ref{fig:dens-tem}). 
This model is also acceptable with $\chi^2/\nu = 248/262$.
The obtained slope of the total mass, $\eta= 1.53$, is consistent 
with the result from the brightness profile analysis (\S\ref{sect:radial}, Figure~\ref{fig:nfw'-mass}), 
where we assumed the isothermality of the ICM.
In fact, the temperature in this solution is nearly isothermal (Figure~\ref{fig:dens-tem}).
This is consistent with the result based on the spectral analysis (\S\ref{sect:t-m-profiles}).
The acceptable fits (99\% confidence) were obtained with $1.42 < \eta < 1.65$.
When $\eta$ was fixed to be 1.53 and $r_{\rm s}$ was left to be free, 
the acceptable fits were obtained with $r_{\rm s} > 43'$.

Moore et al. (1998) found a density profile steeper than the NFW model, 
through their $N-$body simulations  in a standard CDM cosmology with much higher resolution than NFW97.
They employed the form 
\begin{eqnarray}
\rho _{\rm tot}^{\rm Moore} & \propto  & \left(\frac{r}{r_{\rm s}}\right)^{-1.4} \left[1+\left(\frac{r}{r_{\rm s}}\right)^{1.4}\right]^{-1} \label{eq:moore}
\end{eqnarray}
with $r_{\rm s}$ being 0.18 times the virial radius.
We examined this density model ($r_{\rm s} = 16'$ in the case of A1060) and found that this is also consistent with the {\it ASCA} and {\it ROSAT} data.
This is reasonable, since eq.(\ref{eq:moore}) is fairly close to the best-fit solution based on the power-law model.

We may explain how the total mass density models for $\eta$ outside the above range are rejected by the data.
For example, if $\eta$ is too large,
the model implies a too much mass in the central region, 
requiring a steep ICM pressure gradient in the center 
to balance the deeper gravitational potential.
Since the ICM density profile is tightly constrained by the PSPC brightness data, 
the ICM temperature has to be higher towards the center.
This temperature increase becomes inconsistent with the GIS spectrum in the inner region.
Similarly, too small values of $\eta$ require a too much central temperature decrease to meet the GIS spectra.

\subsubsection{The King-type Potential Model}
We also examine a more traditional King-type total mass profile of 
\begin{eqnarray}
\rho _{\rm tot}^{\rm King} & \propto & 
\left[1+\left(\frac{r}{r_{\rm c}}\right)^2\right]^{-\frac{3}{2}}, \label{eq:king-density}
\end{eqnarray}
having a flat density core at the center.
This is an approximation to the inner portions of a self-gravitating sphere.
At large radii, the density is proportional to $r^{-3}$, 
as the NFW [eq.(\ref{eq:nfw-dm})] and power-law density [eq.(\ref{eq:nfw-dm2})] models.
For the ICM density, 
we could assume a $\beta$ model as usual.
However, 
we already found in \S\ref{sect:pspc-density} that the isothermal $\beta$ model can not describe the observed PSPC radial brightness, 
regardless of the potential profile.
Therefore, we again employ the modified NFW ICM model [eq.(\ref{eq:md-nfw-gas2})],  
as in the above two cases.

Utilizing the same method as in the previous subsections, 
we have obtained the best-fit model with $r_{\rm c}=5'.8\ (83h_{70}^{-1} {\rm kpc}),\ a=36'.1, B=10.4$, and $n=0.97$ (dotted lines in Figure \ref{fig:dens-tem}).
The fit is however poorer ($\chi^2/\nu = 291/262$) than the NFW or power-law models,
and can be rejected with 90\% confidence.
The king potential, by its nature, has a flat density core at the center, 
while the ICM density has a cusp.
To compensate the density increase and make pressure to balance the potential, 
the temperature at the center is required to drop as seen in Figure ~\ref{fig:dens-tem}.
This disagree with the good isothermality in the central region:
the spectral discrepancy would become even severer if we take into account the SIS spectrum.


\section{SUMMARY AND DISCUSSION}
Using the high quality data from {\it ASCA} and {\it ROSAT}, 
we have constrained the total mass profile in the central region ($14h_{70}^{-1} \sim 300h_{70}^{-1}$ kpc in radius) of A1060.
The spatially-resolved X-ray spectra obtained with the {\it ASCA} SIS, GIS, and {\it ROSAT} PSPC 
have been found to be consistent with the isothermality of the ICM in the central region of A1060.
Hence, for the first step, 
we assumed an isothermal ICM distribution and 
derived the gravitational potential profile based on the PSPC radial brightness profile.
The potential was found to be deeper than the universal dark halo potential (NFW model) proposed by NFW96 and NFW97,
of which the density scales as $r^{-1}$ at the center.
The total mass density profile has a central slope roughly proportional to $r^{-1.5}$.

For the second step, 
we allowed the ICM to deviate from isothermality and 
tried to reproduce the GIS and PSPC data simultaneously
either by the NFW model, its modification (power-law model), or widely-assumed King model.
Among the three,
the first two models have given successful descriptions of the spectral and spatial data.
On the other hand, we could find no acceptable solution based on the King potential with a flat density core.
The king model requires a temperature profile deviated from isothermality, 
resulting in the poorer fit to both the GIS and PSPC data.
Therefore, 
we conclude that the total mass profile and the ICM temperature profile implied by the King model
is less likely than those implied by the NFW or power-law density solution.

\begin{figure}
\centerline{\hbox{
\psfig{figure=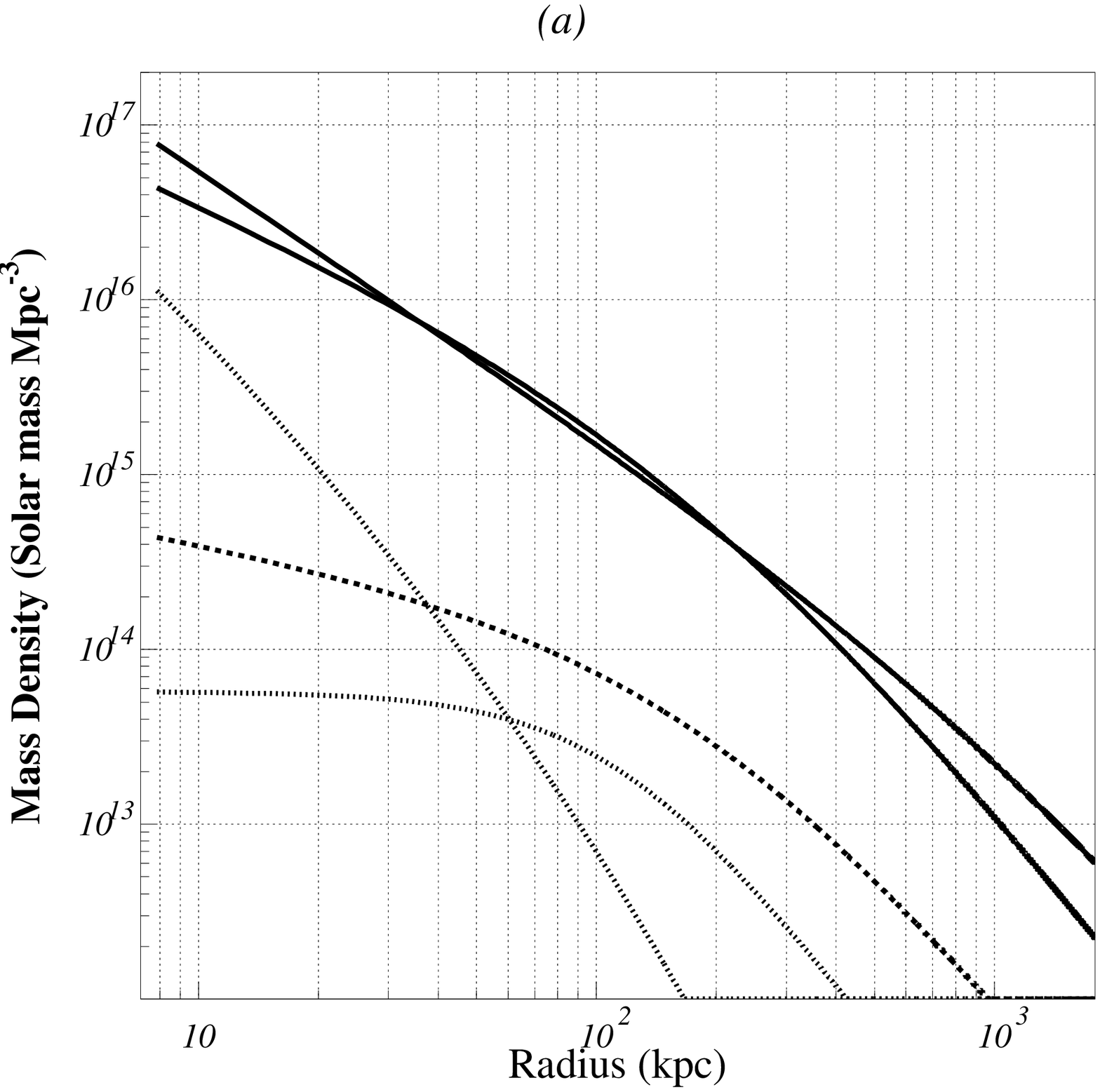,width=0.5\textwidth}
\psfig{figure=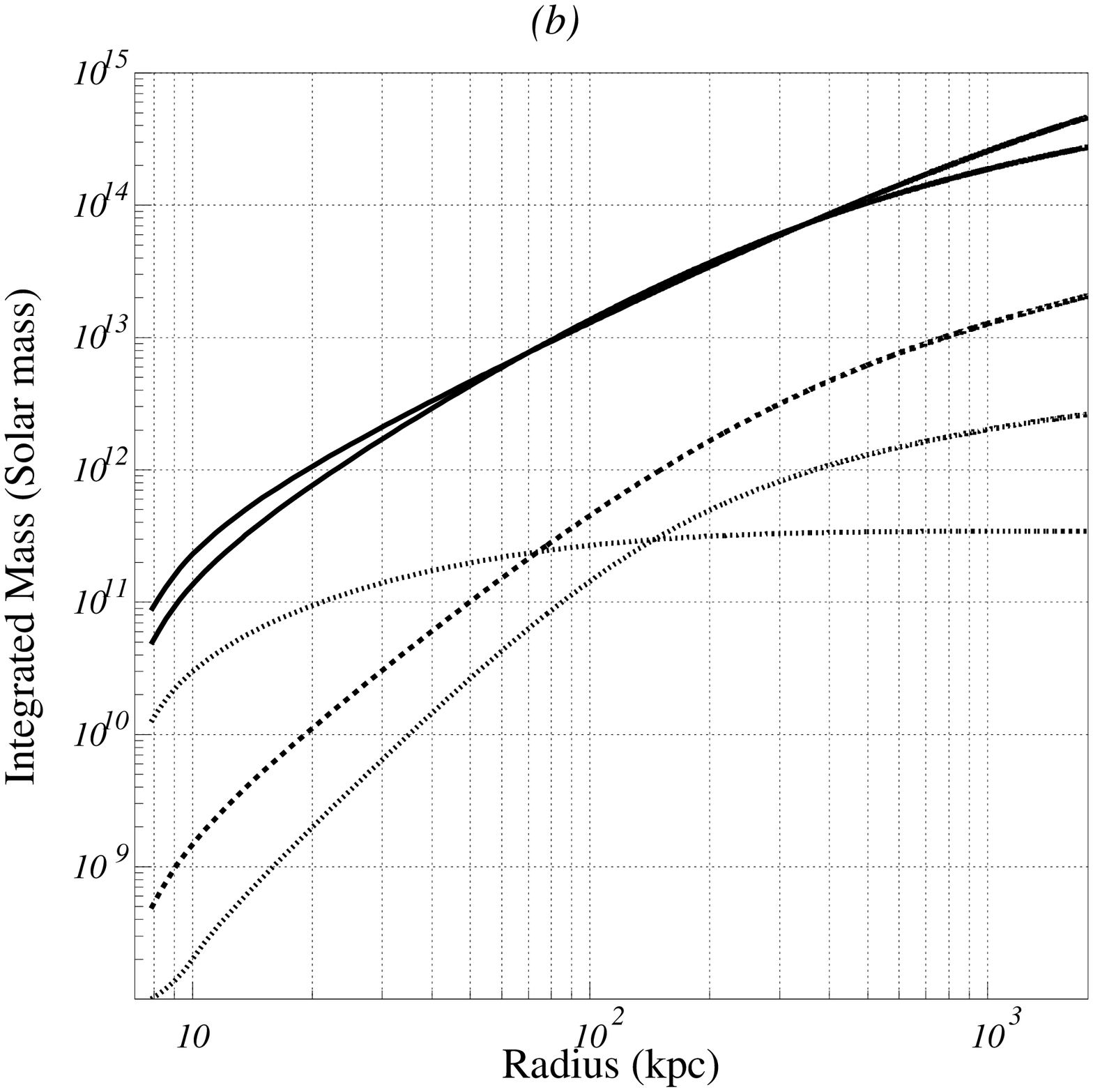,width=0.5\textwidth}
}}
\leavevmode\psfig{figure=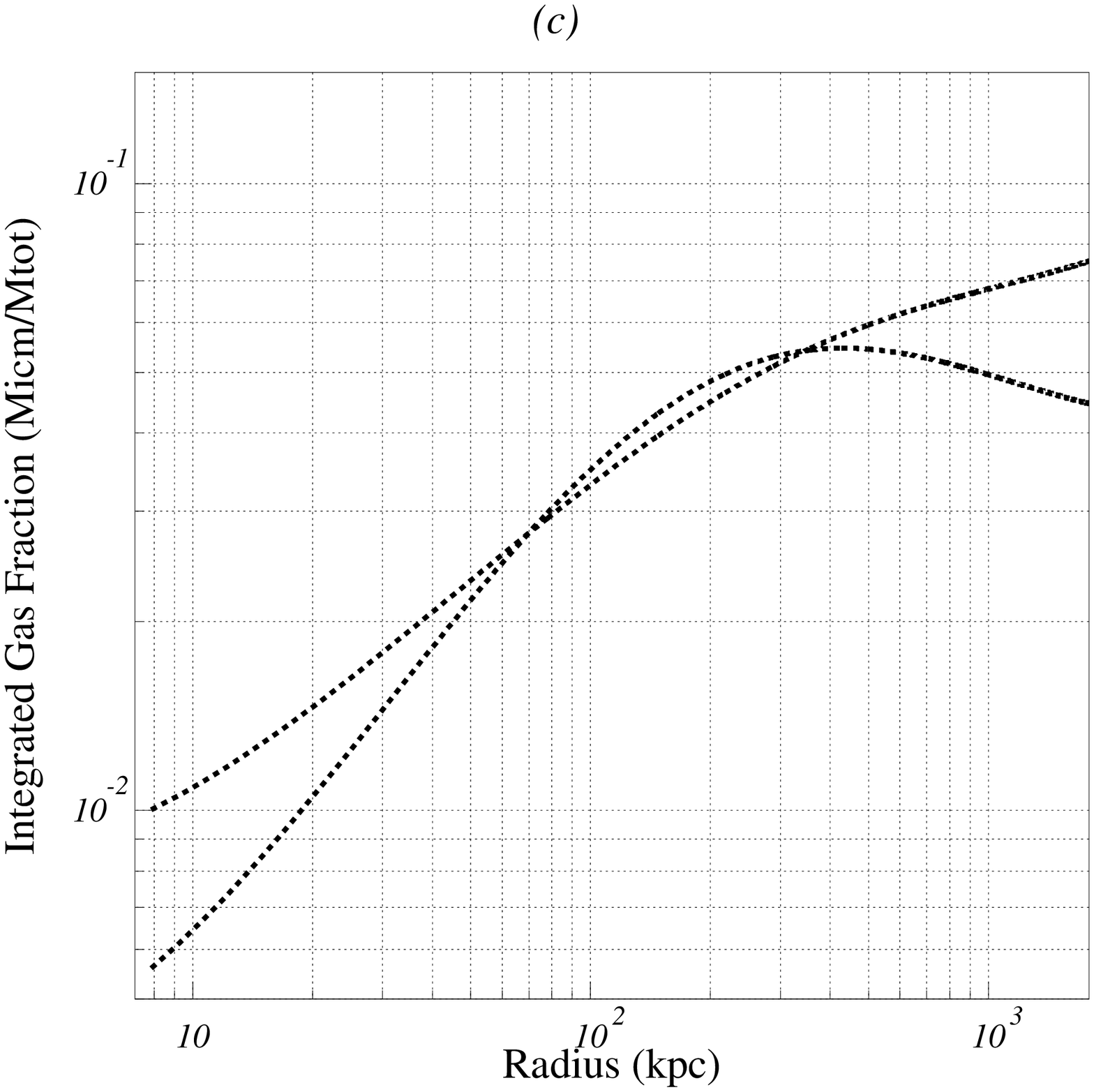,width=0.5\textwidth}
\figcaption{(a) Total mass density profiles from the NFW and power-law density solutions (solid lines). 
For comparison, 
calculated density profiles of the stellar mass  in the central galaxies, 
that in the cluster (dotted lines), 
and the ICM mass (dashed line) are shown. 
(b) The integrated forms of the mass densities.
(c) The integrated ICM fraction relative to the total mass.
We assumed $h_{70}=1.0$.
\label{fig:density-mass}}
\end{figure}

Based on the NFW and power-law density solutions, 
we evaluate the radial density profile of various mass components.
In Figure~\ref{fig:density-mass} (a) and (b) we show the total mass profile, together with
the ICM and stellar mass density profiles.
The ICM density was derived from the X-ray surface brightness distribution, 
while the stellar mass densities were estimated from the optical luminosity distributions of the central galaxy (NGC~3311; V\"asterberg et al. 1991) and that of the cluster (Fitchett and Merrit 1988),
assuming a constant mass-to-light ratio of $M_{\rm stellar}/L_{\rm blue} = 8$, 
where $M_{\rm stellar}$ and $L_{\rm blue}$ represent the mass of the stellar component and the blue luminosity, respectively.
In the region of $14h_{70}^{-1}{\rm kpc}<{\rm r}< 290h_{70}^{-1}{\rm kpc}$, 
the total mass of the luminous matter (the ICM and stellar component) is less than 20\% of the derived total mass,
and hence the dark matter dominates the total mass.
This is consistent with the previous estimate of the baryon fraction in A1060 (Loewenstein and Mushotzky 1996), 
namely 11--16\% within 500 kpc, assuming $H_0 = 50$ km s$^{-1}$Mpc$^{-1}$, 
and justified our formalism in which the baryonic component has been neglected.
Therefore we conclude that 
the dark matter radial profile in A1060 has a central density cusp, 
which is consistent with the NFW prediction, 
instead of a flat density core.

In Figure~\ref{fig:density-mass} (c) we plot the integrated ICM mass fraction to the total mass as a function of radius based on our solutions.
This plot clearly indicate that the total mass and hence the dark matter are much more concentrated than the ICM, 
over the region of $20 h_{70}^{-1} {\rm kpc} \lesssim r \lesssim 200 h_{70}^{-1}{\rm kpc}$, 
or on the ``galaxy scale''.
What produces this distinction?
We consider this to be a result of the difference in temperature distribution between the ``collision-less'' dark matter and collisional ICM.
In a bottom-up scenario of cosmic structure formation,
small scale structure having small amount of gravitational potential collapses in early stage
and settles into the cluster center.
On the other hand, 
the entire cluster as a whole collapses more recently when the total gravitational energy becomes large.
As a result, ``a dynamically cool central region'' develops and remains in the dark matter 
because of a lack of interaction between the particles.
Consequently, the dark matter is concentrated towards the center.
This kind of temperature inversion was predicted through
$N-$body simulations based on the hierarchical clustering universe (e.g., NFW96; Fukushige \& Makino 1997).
In contrast, such a dynamically cool central region did not develop in the ICM,
as a result of a strong heat conduction.
Hence, the ICM is more spread-out than the dark matter.

In addition to the difference in dynamics between the dark matter and ICM, 
extra-gravitational heating could make the ICM more extended than the dark matter.
For example, galactic winds during the cluster formation stage could heat the ICM effectively.
This pre-heating may result in excess entropy of the central ICM.
Actually Ponman, Damian, \& Navarro (1999) found the excess in clusters and galaxy group with ICM temperatures of $<4$ keV, 
by comparing central ICM entropy and ICM temperature of clusters 
with those obtained from $N-$body/gas-dynamical simulations by Eke, Navarro, \& Frenk (1998; hereafter ENK98).
They calculated the entropy, $T/n_{\rm e}^{2/3}$, 
where $T$ and $n_{\rm e}$ are the ICM temperature and electron density, at a fiducial radius of 0.1 times the virial radius.
Note that ENK98 does not include the extra-gravitational heating.
Being similar to samples in 
Ponman, Damian, \& Navarro (1999), 
A1060 has higher entropy by a factor of $\sim 2$ than the prediction by ENK98.
We also note that some studies based on $N-$body/gas-dynamical simulations
indicated that the heating of the ICM before the gravitational collapse strongly affects the evolution of the ICM properties and hence the statistical properties of clusters,
but insignificantly affects the ICM distribution of the equilibrium system (Metzler \& Evrard 1994; Navarro, Frenk, \& White 1995).
Furthermore these simulations and ENK98 do not include radiative cooling of the ICM, which could affects the central distribution of the ICM as well as the pre-heating.
Therefore, we consider that it is rather difficult to find evidence for the extra-gravitational heating by comparing the derived ICM distribution of a cluster 
with those derived theoretically at this time.


Based on the NFW model, 
we found the scale length $r_{\rm s}$ and the normalization of the total mass in unit of the critical density of the universe $\delta_{\rm c}$ to be $(190^{+60}_{-30})h_{70}^{-1}$ kpc and $(1.2^{+0.5}_{-0.4})\times 10^{4}$, respectively.
The virial mass $M_{200}$ (defined e.g., in NFW97) can be calculated as $(2.1^{+0.5}_{-0.2} \times 10^{14})h_{70}^{-1}${\it M$_{\odot}$}.
Through $N-$body simulation based on the standard CDM universe ($\Omega_0 = 1$), 
NFW97 predicted $\delta_{\rm c}$ of $\sim 2\times 10^{4}$ for the virial mass of $2 \times 10^{14}h_{70}^{-1} ${\it M$_{\odot}$} (see Figure 9 in NFW97), 
which agrees with our result within a factor of 2.
In the Centaurus cluster, 
Ikebe et al. (1999) also found similar agreement  between the X-ray measurements and the NFW prediction.

We have shown that the total mass profile having 
a steeper density profile of $\rho \propto r^{-\eta}$ with $1.42<\eta<1.65$ (the power-law solution) in the central region
is consistent with the data, 
as is the original NFW model of $\rho \propto r^{-1}$.
Recent $N-$body simulations with much higher resolution than those of NFW97 show such steeper density profiles of dark halos (e.g. Fukushige \& Makino 1997; Moore et~ al. 1998), in a nice agreement with our best-fit solution.
In fact, we showed that a density profile found by Moore et~ al. (1998) through the $N-$body simulation with the highest resolution is consistent with the data.
Accordingly, results of $N-$body simulations in hierarchical clustering universe 
are consistent with our observational results.

\acknowledgments
We thank all the members of the ASCA team led by Y. Tanaka and the ROSAT team for making this study possible.
We also thank S. Sasaki, Y. Suto, M. Sekiguchi, and H. B\"ohringer for their constructive discussion.
We appreciate the referee who provided valuable comments.
ASCA data were mainly analized utilizing software developed by the ASCA-ANL and SimASCA team.
The ROSAT data were obtained through ROSAT Archive Browser 
, provided by Max-Planck-Institut f\"ur extraterrestrische Physik.
T.T. is supported by the post-doctoral program of Japan Society
for the Promotion of Science.

\newpage









\begin{deluxetable}{ccccccc}
\tablecaption{Single temperature fits to the spectra within the central $5'$.\label{tbl:r5spectra}}
\tablewidth{0pt}
\tablehead{
\colhead{Detector}	& \colhead{$N_{\rm H}$}	& \colhead{$kT$} & \colhead{Metallicity}&  \colhead{$\chi ^2/\nu$}\\
\colhead{}		& \colhead{($10^{20}$cm$^{-2}$)}& \colhead{(keV)} & \colhead{(Solar)}&  \colhead{}
}
\startdata
GIS\tablenotemark{a}	& 3.5 (1.3--5.7)		& 3.3 (3.2--3.4)	& 0.31 (0.25--0.37) & 122/110	\nl
SIS\tablenotemark{b}	& 7.9 (7.4--8.7)		& 3.3 (3.2--3.4)	& 0.28 (0.26--0.32) & 374/309	\nl
PSPC\tablenotemark{c} 	& 6.1 (5.3--6.4)		& 2.2 (2.1--2.8)	&  0.32 (0.24--0.50) & 45/37	\nl
joint\tablenotemark{d}		& \nodata		& 3.3 (3.2--3.3)	& 0.29 (0.27--0.33)  &556/459	\nl
\enddata

\tablenotetext{a}{In the 0.7--10.0 keV energy band.}
\tablenotetext{b}{In the 0.5--8.0 keV energy band.}
\tablenotetext{c}{In the 0.3--2.0 keV energy band.}
\tablenotetext{d}{The normalization and column density were allowed to take independent values among three detectors.}
\end{deluxetable}

\begin{deluxetable}{cccccc}
\tablecaption{Results of the $\beta$ model, NFW ICM, and modified NFW ICM model fits to the GIS and PSPC radial count-rate profiles of A1060.\label{tbl:rpfit}}
\tablewidth{0pt}
\tablehead{
\colhead{Model}	& \colhead{Detector}	& \colhead{$r_{\rm c}$ or $r_{\rm s}$} & \colhead{$\beta$ or $B$}	& \colhead{$n$}	& \colhead{$\chi ^2/\nu $} \\
\colhead{}	& \colhead{}		& \colhead{(arcmin.)}		& \colhead{}	& \colhead{}	& \colhead{} 
}
\startdata
$\beta$ model	& GIS\tablenotemark{a}	& 4.2 (3.6--4.8) & 0.57 (0.54--0.61) && 9.48/17\nl
		& PSPC\tablenotemark{b}	& 3.9 (3.8--4.0)  & 0.54 (0.53--0.55) && 264/96\nl
NFW\tablenotemark{c}	& GIS & 12.4 (11.0--14.0)	& 7.8 (7.4--8.2)	&& 11.2/17\nl
		& PSPC	& 13.8 (13.3--14.3)  & 8.0 (7.8--8.2) && 132/96\nl
md-NFW\tablenotemark{d} & PSPC &  41.8 (40--43)		& 11.0 (10.7--11.3)	& 0.972 (0.969--0.975)& 69/95\nl

\enddata
\tablenotetext{a}{The GIS covers 0.7--10.0 keV in energy and up to $20'$ in radius.}
\tablenotetext{b}{The PSPC covers 0.5--2.0 keV in energy and $50'$ in radius.}
\tablenotetext{c}{The NFW ICM model.}
\tablenotetext{d}{The modified NFW ICM model.}
\end{deluxetable}

\clearpage

\end{document}